\documentstyle[multicol,epsf,aps]{revtex}
\begin{document}
\title{Ballistic Annihilation}
\author{Paul L. Kaprivsky$^{1,2}$ and Cl\'ement Sire$^1$}
\address{$^1$ Laboratoire de Physique Quantique (UMR C5626 du CNRS),
Universit\'e Paul Sabatier, 31062 Toulouse Cedex, France \\ $^2$
Center for Polymer Physics and Department of Physics, Boston
University, Boston, MA 02215, USA}
\maketitle
 
\begin{abstract} 

\noindent
Ballistic annihilation with continuous initial velocity distributions
is investigated in the framework of Boltzmann equation. The particle
density and the rms velocity decay as $c\sim t^{-\alpha}$ and $\langle
v\rangle\sim t^{-\beta}$, with the exponents depending on the initial
velocity distribution and the spatial dimension. For instance, in one
dimension for the uniform initial velocity distribution we find
$\beta=0.230472\ldots$. We also solve the Boltzmann equation
for Maxwell particles and very hard particles in arbitrary spatial
dimension. These solvable cases provide bounds for the decay exponents
of the hard sphere gas.

\smallskip\noindent{PACS numbers: 05.20.Dd, 03.20.+i, 82.20Mj}
\end{abstract}

\begin{multicols}{2} 

Ballistic annihilation is the kinetic process which involves particles 
undergoing ballistic motion and annihilating upon colliding.  Ballistic 
annihilation underlies numerous apparently unrelated phenomena, e.g., 
growth and coarsening processes and traffic flows\cite{ks,sek,fkb,gk}.
In coarsening processes, for instance,  domain walls have a natural 
particle interpretation.  Despite all that, little is known on
irreversible processes where the reactants move ballistically while 
the contrasting situation of diffusion-controlled processes is well 
understood\cite{privman}. 

There exist few theoretical results on ballistic annihilation in one
dimension with {\em discrete} velocity distributions. For the simplest
binary velocity distribution, the ballistic annihilation process has
been solved in one dimension by Elskens and Frisch\cite{els}, see also
Refs.\cite{ks,bf,fk,more}; some analytical results are also available
for the ternary velocity distribution \cite{krl,Piasecki}.  No
solutions have been found for {\em continuous} initial velocity
distributions, although the decay exponents have been determined
numerically\cite{brl,rdp}. This lack of analytical results is
especially striking given that ballistic aggregation processes admit
exact solutions in one dimension for arbitrary initial velocity
distributions\cite{bkr,laur}.

In this work, we consider ballistic annihilation with continuous
isotropic initial velocity distributions in arbitrary dimension.  Our
analysis is performed in the framework of Boltzmann equation approach.
This scheme involves an uncontrolled approximation and generally leads
to erroneous results for ballistic annihilation with discrete velocity
distributions.  For continuous velocity distributions, however, the
decay exponents found from numerical integration of the Boltzmann
equation\cite{brl} are an excellent agreement with simulation
results\cite{brl,rdp}. Furthermore, for solvable one dimension
ballistic aggregation processes with continuous velocity
distributions\cite{bkr,laur}, the decay exponents computed from the
Boltzmann equation are exact; for higher-dimensional ballistic
aggregation process\cite{cpy}, the decay exponents determined within
the Boltzmann framework\cite{cpy,pe} appear to be exact as well.  
Therefore, we believe that the Boltzmann equation approach provides
exact decay exponents for ballistic annihilation with continuous
velocity distributions.

For clarity, we start with one-dimensional ballistic annihilation
process.  An appropriate Boltzmann equation reads\cite{brl}
\begin{equation}
\label{be}
{\partial P(v,t)\over \partial t}=
-P(v,t)\int_{-\infty}^\infty dv' |v-v'|P(v',t).
\end{equation}
In the long time limit, the velocity distribution approaches a scaling 
form
\begin{equation}
\label{scal}
P(v,t)=t^{\beta-\alpha}F(x), \quad {\rm with}\quad x=vt^\beta.
\end{equation}
The two basic exponents determine the behavior of 
the particle concentration $c(t)$ and the rms velocity $\langle v\rangle$:
\begin{equation}
\label{expdef}
c(t)\sim t^{-\alpha}, \quad 
\langle v\rangle \sim  t^{-\beta}.
\end{equation}
Formally, $c$ and $\langle v\rangle$ are defined as follows
\begin{eqnarray*}
c(t)&=&\int_{-\infty}^\infty dv\,P(v,t)
=t^{-\alpha}\int_{-\infty}^\infty dx\,F(x),\\
\langle v\rangle^2 &=&c^{-1}\int_{-\infty}^\infty dv\,v^2 P(v,t)
=t^{-2\beta}{\int_{-\infty}^\infty dx\,x^2F(x)\over 
\int_{-\infty}^\infty dx\,F(x)}.
\end{eqnarray*}

By inserting the scaling form (\ref{scal}) into Eq.~(\ref{be}) one finds
$\alpha+\beta=1$ and an equation for the scaling function
\begin{equation}
\label{Fx}
2\beta-1+\beta x\,{F'(x)\over F(x)}
=-\int_{-\infty}^\infty dx'\,|x-x'|\,F(x').
\end{equation}
In the following, we always consider isotropic initial velocity
distributions.  In one dimension, this requirement reads
$P_0(v)=P_0(-v)$ and it implies the symmetry for later times
$P(v,t)=P(-v,t)$ and the symmetry of the scaling function
$F(x)=F(-x)$.  

The large $x$ behavior of the scaled velocity distribution is found by
noting that in this region the integral on the right-hand side of
Eq.~(\ref{Fx}) simplifies to $Cx$, where $C$ is the normalization
constant, $C=\int_{-\infty}^\infty dx\,F(x)$.  Solving the resulting
differential equation gives
\begin{equation}
\label{inf}
F(x)\sim x^{(1-2\beta)/\beta}\,e^{-Cx/\beta}
\quad {\rm when} \quad x\to\infty.
\end{equation}

For a given $\beta$, a solution of Eq.~(\ref{Fx}) would not agree with
the boundary condition at $x=0$ which is implied by the initial
velocity distribution. Thus, we arrive at an eigenvalue problem.  For
instance, we should require $F'(0)=0$ if the initial velocity
distribution is flat near the origin. Solving this eigenvalue problem
numerically gives $\beta_{\rm flat}=0.230472\ldots$, to be compared
with $\beta_{\rm flat}\approx 0.22$\cite{brl} and $\beta_{\rm
flat}\approx 0.19$\cite{rdp} found from Monte Carlo simulations.
Similarly, for the initial velocity distribution satisfying
$P_0(v)\sim |v|$ we find $\beta_{\rm linear}=0.166649\ldots$.

We now outline an approximate analytical computation of the exponent
$\beta$.  First we note that Eq.~(\ref{Fx}) can be reduced to an
ordinary differential equation after a double differentiation. This
equation can be further simplified by the rescaling $x\to
x\sqrt{\beta}$ that eliminates the $\beta$ factor. The governing
equation then reads
\begin{equation}
\label{diff}
 xf'''+2f''=2\exp(-f),
\end{equation}
with $f(x)=-\ln F(x)$.  We shall also use the relation
\begin{equation}
\label{norm2}
\int_0^\infty dx\,x\,F(x)=(2\beta)^{-1}-1,
\end{equation}
which will play the role of a normalization condition. Equation
(\ref{norm2}) is just Eq.~(\ref{Fx}) at $x=0$. 

Although Eq.~(\ref{diff}) cannot be solved exactly, an approximate
solution $f_1$ can be found by replacing the right-hand side by
$e^{-f_0}$, where $f_0$ is a reasonable approximation for $f$. In
principle, this approximation scheme can be repeated again starting
from $f_1$, and should finally lead to the exact form for
$f$.  As the starting point, we choose
\begin{equation}
\label{F0}
F_0(x)=e^{-f_0(x)}=\left(1+{x\over \nu}+{1-\nu^2\over 2\nu^2}\,x^2\right)
\,e^{-x/\nu}.
\end{equation}
The function $F_0(x)$ is constructed in such a way, that the small $x$
expansion is correct up to the second order.  Additionally, $F_0(x)$
exhibits an exponential decay for large $x$, in agreement with the
exact asymptotic behavior (\ref{inf}).  The parameter $\nu$ is yet to
be determined.  Replacing now the right-hand side of Eq.~(\ref{diff})
by $F_0(x)$ from Eq.~(\ref{F0}) and solving the resulting {\em linear}
differential equation gives
\begin{eqnarray}
\label{happ}
f_1(x)&=&2\nu(3-\nu^2)x+\nu^2(7-5\nu^2)\left(1-e^{-x/\nu}\right)\nonumber\\
&-&\nu(1-\nu^2)x\,e^{-x/\nu}\nonumber\\
&-&6\nu^2(2-\nu^2)\int_0^{x/\nu}\frac{1-\exp(-t)}{t}\,dt.
\end{eqnarray}
The constant $\nu$ can now be calculated self-consistently by imposing 
the constraint
\begin{equation}
\label{scc}
\nu^{-1}=2\int_{0}^\infty F(x)\,dx,
\end{equation}
which can be obtained by integrating Eq.~(\ref{diff}).  Plugging
$F_{\rm appr}=e^{-f_1(x)}$ into Eq.~(\ref{scc}) gives $\nu^{-1}_{\rm
appr}\cong 2.67156$ to be compared to the exact numerical value
$\nu^{-1}=2.65826\ldots$. Finally, Eq.~(\ref{norm2}) leads to
$\beta_{\rm appr}\cong 0.22898$, in good agreement with the exact
value obtained above. In Fig.~1, we plot the exact numerical scaling
function $F$ and the approximate solution $F_{\rm appr}=e^{-f_1(x)}$
with $f_1(x)$ given by Eq.~(\ref{happ}). The approximate solution is
extremely good as the relative error is always less than 0.5\%.

\begin{figure}[h]
\narrowtext      
\epsfxsize=\hsize
\epsfbox{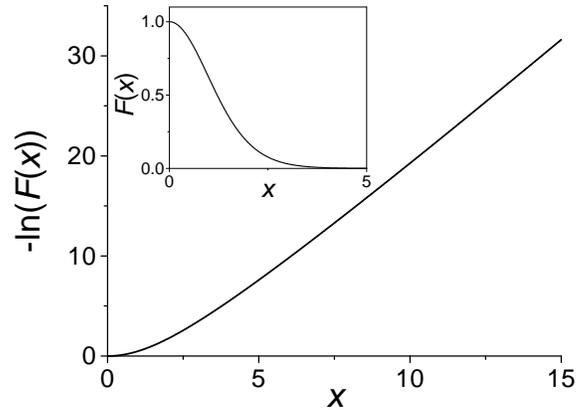} 
\vskip 0.5cm
\caption{Plot on a semi-log 
scale of the scaling function $F$ and its approximation given by
Eq.~(\ref{happ}). The insert shows both functions in normal
scale. Both scaling functions are indistinguishable as the relative
error is less than 0.5\%}
\end{figure}   
 
In higher dimensions, the Boltzmann equation reads
\begin{equation}
\label{BE}
{\partial P({\bf v},t)\over \partial t}=
-P({\bf v},t)\int d{\bf w}\, |{\bf v}-{\bf w}|P({\bf w},t).
\end{equation}
For isotropic initial velocity distributions, the appropriate scaling 
variable is $x=vt^\beta$ with $v=|{\bf v}|$, and the scaling form is
$P({\bf v},t)=t^{d\beta-\alpha}F(x)$.  Plugging this scaling form into
Eq.~(\ref{BE}) and using the spherical coordinates to simplify the 
collision integral we obtain
\begin{eqnarray}
\label{dFx}
(d+1)\beta&-&1+\beta x\,{F'(x)\over F(x)}\\
&=&-\Omega_{d-1}\int_0^\infty dy\,y^{d-1}\,E_d(x,y)\,F(y).\nonumber
\end{eqnarray}
Here $\Omega_{d-1}$ is the surface area of the unit sphere in $d-1$ dimensions
and 
\begin{equation}
\label{E}
E_d(x,y)=\int_0^\pi d\phi\, (\sin \phi)^{d-2}\sqrt{x^2+y^2-2xy\cos\phi}.
\end{equation}
In two dimensions, $E_2(x,y)=2(x+y)E(k)$, where $E(k)$ is the complete 
elliptic integral of the second kind with modulus $k={2\sqrt{xy}\over x+y}$. 
Similarly, in an arbitrary even dimension $E_d(x,y)$ can be expressed via 
elliptic integrals. In odd dimensions, $E_d(x,y)$ can be expressed in 
terms of elementary functions. In the most interesting three dimensional 
case, one reduces Eq.~(\ref{dFx}) to
\begin{eqnarray}
\label{3Fx}
4\beta&-&1+\beta x\,{F'(x)\over F(x)}\\
&=&-{2\pi\over 3}\int_0^\infty dy\,y\,
{(x+y)^3-|x-y|^3\over x}\,F(y).\nonumber
\end{eqnarray}
One should solve Eq.~(\ref{dFx}) subject to an appropriate boundary
condition at the origin. Overall, the task reduces again to the
eigenvalue problem.  Note that the approximation scheme explicitly
presented above in $d=1$ could be equally applied in higher
dimensions.

Thus, the decay exponents can be determined with arbitrarily high
precision in arbitrary spatial dimension for arbitrary isotropic
initial velocity distribution. This method, however, does not solve
the Boltzmann equation (\ref{be}). We now provide two {\em
approximate} solutions to the Boltzmann equation for arbitrary initial
conditions and arbitrary spatial dimension $d$.  These
solutions are in fact {\em exact} solutions of Boltzmann equations with
collision kernels $\sigma(g)$ different from $\sigma(g)=g$
characterizing the hard sphere gas (here $g\equiv |{\bf v}-{\bf w}|$
is the relative velocity). Note that such collision kernels naturally
arise in kinetic theory of interacting particles\cite{r}.  In the
present context, the generalized Boltzmann equation reads
\begin{equation}
\label{GBE}
{\partial P({\bf v},t)\over \partial t}=
-P({\bf v},t)\int d{\bf w}\,\sigma(g)\,P({\bf w},t).
\end{equation}
Let us compare dimensions of the left and right-hand side of
Eq.~(\ref{GBE}).  The velocity distribution has dimension
$[P]=T/L^{d+1}$ which implies $[\sigma]=L^d/T$; therefore
$[\sigma(g)/g]=L^{d-1}$.  The remaining quantity with dimension of
length, the ``interaction'' radius, should be extracted from the
collision process.  For hard sphere gas, the relevant interaction
radius is simply the geometrical radius $a$ of the spheres, so
$\sigma(g)=g a^{d-1}$. The constant factor $a^{d-1}$ can be absorbed
into the time variable -- this is what we have done in Eq.~(\ref{BE}).
For particles interacting through a two-body power law potential,
$U(r)\propto r^{-n}$, the energy conservation implies $g^2\sim
r^{-n}$. Thus, $\sigma(g)\sim g r^{d-1}\sim g^\nu$ with
$\nu=1-{2(d-1)\over n}$.  The hard sphere gas ($\nu=1$) is recovered
for $n=\infty$.  The velocity independent kernel ($\nu=0$), the
so-called ``Maxwell'' gas, arises when particles interact through the
power law potential with the exponent $n=2(d-1)$.  When $\nu>1$, the
interaction is ``harder'' than in the hard sphere gas (though such
behavior does not arise from a simple power-law interaction
potential). One particularly tractable model corresponds to
$\sigma(g)=g^2$, the so-called gas of very hard particles\cite{ernst}.

To provide a faithful analog of the original hard sphere gas, we
replace $\sigma(g)=g$ by $\langle v\rangle$ in the Maxwell case and by
the factor $g^2/\langle v\rangle$ for the very hard particles.  Hence
for the Maxwell gas the Boltzmann equation (\ref{GBE}) becomes
\begin{equation}
\label{ME}
{\partial P({\bf v},t)\over \partial t}
=-P({\bf v},t) c(t)\langle v(t)\rangle.
\end{equation}
Thus we get effectively non-interacting particles as different velocities
remain uncoupled. Solving (\ref{ME}) yields
\begin{equation}
\label{MEsol}
P({\bf v},t)=c(t)P_0({\bf v}), \quad
c(t)={1\over 1+\langle v\rangle t}.
\end{equation}
The moments of the velocity do not change with time, e.g., $\langle
v(t)\rangle=\langle v\rangle_0$, and thence $\alpha=1$ and
$\beta=0$. 

More interesting results are found for the very hard particles.  The
corresponding Boltzmann equation, i.e., Eq.~(\ref{GBE}) with 
$\sigma(g)=g^2/\langle v\rangle$, can be simplified by absorbing the 
$\langle v\rangle^{-1}$ factor into the time variable,
\begin{equation}
\label{tau}
\tau=\int_0^t{dt'\over \langle v(t')\rangle},
\end{equation} 
and reducing the collision integral into a combination 
of the moments
$M_j(\tau)=\int d{\bf w}\,w^j P({\bf w},\tau)$ of the velocity distribution.
The Boltzmann equation becomes 
\begin{equation}
\label{vhp}
{\partial P({\bf v},\tau)\over \partial \tau}=-P({\bf v},\tau)
\left[v^2 M_0(\tau)+M_2(\tau)\right],
\end{equation}
and easily solved to give
\begin{equation}
\label{solvhp}
P({\bf v},\tau)=P_0({\bf v})\,e^{-v^2 L_0(\tau)-L_2(\tau)}.
\end{equation}
Here $L_j(\tau)=\int_0^\tau d\tau'\,M_j(\tau')$. 

To derive explicit results, it is natural to consider initial velocity
distributions algebraic near the origin.  To simplify algebra, we
specifically choose
\begin{equation}
\label{initialD}
P_0({\bf v})={2\,v^{\mu}\,e^{-v^2}\over 
\Omega_d\Gamma\left[(\mu+d)/2\right]},
\end{equation}
where $\Omega_d=2\pi^{d/2}/\Gamma(d/2)$ is the surface area of the
unit sphere in $d$ dimensions, and the prefactor in (\ref{initialD})
is chosen to set the initial density to unity. Combining
(\ref{solvhp}) and (\ref{initialD}) we can explicitly compute $M_0$
and $M_2$. Using then $M_j={dL_j\over d\tau}$ closes the problem. We
find
\begin{eqnarray*}
{dL_0\over d\tau}={e^{-L_2}\over (1+L_0)^{\mu+d\over 2}}, \quad
{dL_2\over d\tau}={\mu+d\over 2}\,{e^{-L_2}\over (1+L_0)^{\mu+d+2\over 2}}.
\end{eqnarray*}
Solving these equations subject to $L_0(0)=L_2(0)=0$ yields
\begin{eqnarray*}
L_0&=&\left[1+(\mu+d+1)\tau\right]^{1\over \mu+d+1}-1, \\
L_2&=&{\mu+d\over 2(\mu+d+1)}\,\, \ln\left[1+(\mu+d+1)\tau\right].
\end{eqnarray*}
Now we compute $c=M_0={dL_0\over d\tau}$ to find the density,
\begin{equation}
\label{cvhp}
c=\left[1+(\mu+d+1)\tau\right]^{-{\mu+d\over \mu+d+1}}, 
\end{equation}
and $\langle v\rangle^2={M_2\over M_0}={\mu+d\over2}(1+L_0)^{-1}$,
\begin{equation}
\label{vvhp}
\langle v\rangle=\sqrt{\mu+d\over 2}
\left[1+(\mu+d+1)\tau\right]^{-{1\over 2(\mu+d+1)}}.
\end{equation}
By inserting Eq.~(\ref{vvhp}) into Eq.~(\ref{tau}) we can express $\tau$
via the original time variable $t$. Finally, we arrive at
\begin{equation}
\label{expvhp}
\alpha={2(\mu+d)\over 2(\mu+d)+1}, \quad 
\beta ={1\over 2(\mu+d)+1}.
\end{equation}
Note that the exponent relation $\alpha+\beta=1$ which is valid for any 
$\mu$ and $d$. This sum rule immediately follows from an elementary 
mean-free path argument: $a^{d-1}c\langle v\rangle t\sim 1$.

The above exact values of the exponents in the two solvable limits appear 
to provide the strict bounds for the hard sphere case:
\begin{equation}
\label{expbounds}
{2(\mu+d)\over 2(\mu+d)+1}<\alpha<1,\quad
0<\beta<{1\over 2(\mu+d)+1}.
\end{equation}
These bounds are fair for small $d$ and they get more and more
stringent as the spatial dimension increases. 

The above method of solving the Boltzmann equation can be adapted to
the more general collision kernels $\sigma({\bf v},{\bf v'})=\langle
v\rangle^{1-\kappa} \left(|{\bf v}|^\kappa+|{\bf v'}|^\kappa\right)$
(the most natural case corresponds to $\kappa=1$).  For $\kappa>0$,
the decay exponents are given by Eqs.~(\ref{expvhp}).  The collision
kernels $\sigma(g)=g^{2n}$ constitute a more perspective
generalization.  One could try to solve the Boltzmann equation when
$n$ is integer and then perform an analytic continuation to $n=1/2$
corresponding to the hard sphere gas.  Besides the cases of Maxwell
and very hard particles ($n=0$ and 1, respectively), it is possible to
work out the case of $n=2$.  Unfortunately, we have not succeeded
beyond that.

In the limit $d\to\infty$, the Boltzmann equation (\ref{BE}) becomes
tractable.  First of all, the collision kernel simplifies to
$\sqrt{v^2+w^2}$ as (different) vectors are orthogonal in infinite
dimensions. The scaled Boltzmann equation reads
\begin{equation}
\label{Finf}
1-B-\beta x\,{F'\over F}
=\Omega_d\int_0^\infty dy\,y^{d-1}\sqrt{x^2+y^2}\,F(y),
\end{equation}
with $B=(d+1)\beta$. The bounds of Eq.~(\ref{expbounds}) lead to
$\beta\to 0$ but remain non-trivial for $B$, $0<B<1/2$.  The
right-hand side of Eq.~(\ref{Finf}) is computed by the saddle point
technique to find $C\sqrt{x^2+y_*^2}$, where $C$ is the normalization
factor $C=\Omega_d\int_0^\infty dy\,y^{d-1}\,F(y)$, and $y_*$ is the
saddle point which is found from $y_*\,F'(y_*)/F(y_*)+d-1=0$.  Hence, at
the saddle point Eq.~(\ref{Finf}) gives $1=Cy_*\sqrt{2}$.  Near the
origin, $F(x)\sim x^\mu$ and thus $\beta x F'/F=\beta\mu\to 0$, so we
find $1-B=Cy_*=1/\sqrt{2}$. Thus we arrive at the universal
asymptotics,
\begin{equation}
\label{expbeta}
\beta\simeq \left(1-{1\over \sqrt{2}}\right)\,d^{-1}
\quad{\rm when}\quad d\to\infty,
\end{equation}
independently on $\mu$.

In this work, we reduced the determination of the decay exponents for
ballistic annihilation to an eigenvalue problem. We found that the
exponents have very non-trivial values even for the simplest initial
velocity distributions.  Our approach manifestly demonstrates that the
decay exponents are affected only by the spatial dimension $d$ and by
the exponent $\mu$ characterizing the initial velocity distribution in
the $|{\bf v}|\to 0$ limit: $P_0({\bf v})\sim |{\bf v}|^\mu$.  We also
solved the Boltzmann equation for the Maxwell particles and very hard
particles in arbitrary spatial dimension.  For the hard sphere gas, we
found the asymptotic behavior of the exponents in the $d\to\infty$
limit.

\medskip
\noindent
Interesting discussions with Eli Ben-Naim and Satya Majumdar are
gratefully acknowledged.  PLK also acknowledges support from NSF,
ARO, and CNRS.

\end{multicols} 
\end{document}